# Observation of Possible Nonlinear Anomalous Hall Effect in Organic Two-Dimensional Dirac Fermion System

Andhika Kiswandhi* and Toshihito Osada**

*Institute for Solid State Physics, The University of Tokyo,*

*5-1-5 Kashiwanoha, Kashiwa, Chiba 277-8581, Japan.*

We report the observation of nonlinear anomalous Hall effect (NLAHE) in the multilayered organic conductor $\alpha$-(BEDT-TTF)$_2$I$_3$ in the charge order (CO) insulating phase just under the critical pressure for transition into two-dimensional (2D) massless Dirac fermion (DF) phase. We successfully extracted the finite nonlinear Hall voltage proportional to square current at zero magnetic field. The observed NLAHE features, current direction dependence and correlation with CO, are consistent with the previous estimation assuming 2D massive DF with a pair of tilted Dirac cones. This is the first observation of topological transport in organic conductors, and also the first example of NLAHE in the electronic phase with spontaneous symmetry breaking.

# 1. Introduction

Nonlinear anomalous Hall effect (NLAHE) in zero magnetic field condition has attracted much interest since its prediction by Sodemann and Fu [1]. The phenomenon is induced by current flowing through the sample and originating from the nonequilibrium distribution of electrons (current-carrying state) on the integral of the Berry curvature. The model is assumed to be time-reversal invariant having Kramers pair states at general **k** and −**k** points in the momentum space, whose Berry curvatures are related by $\Omega(\mathbf{k}) = -\Omega(-\mathbf{k})$, and requires broken inversion symmetry for nonzero Berry curvature. In such system, the Berry curvature summed over the occupied states must vanish in equilibrium condition, so anomalous velocity effectively cancels out. However, in nonequilibrium state by applying electric current, the electron distribution is shifted by Δ**k** in the **k**-space. Therefore, the Kramers pair states at **k** and −**k** are not equally occupied, and the sum now can be finite. Generally, in the system with the time reversal symmetry, the sum of the Berry curvature over occupied states in the nonequilibrium current-carrying state is given by the inner product of Δ**k** and the Berry curvature dipole of the equilibrium state. Due to the finite Berry curvature, the electrons gain a net anomalous velocity $v_\perp$ perpendicular to the applied electric field, similar to Hall effect even in the absence of magnetic field. The anomalous current density, $j_\perp$ is written as $j_\perp = -e \int f(\epsilon_k) v_\perp(\mathbf{k}) d\mathbf{k}$, where $f(\epsilon_k) = f_0(\epsilon_k) + \delta f(\epsilon_k)$ is the nonequilibrium Boltzmann distribution function. Here, $f_0(\epsilon_k)$ is the distribution at equilibrium and $\delta f(\epsilon_k)$ is the correction to the first order in applied electric field $E$. Since the anomalous velocity is also proportional to $E$, the anomalous current density is proportional to $E^2$. Therefore, unlike the conventional Hall effect, the anomalous current exhibits a second-harmonic behavior, implying a rectification effect since its direction is unchanged even when the excitation current direction is reversed. The signal strength depends on the magnitude of the Berry curvature dipole, which is more prominent in systems with tilted Dirac or Weyl cones with a finite gap.

NLAHE has been observed in several materials, such as few-layer 3D Weyl fermion system $WTe_2$ [2,3], bulk 3D Dirac semimetal $Cd_3As_2$ [4], and thin-film $TaIrTe_4$ [5]. Notably, in the latter, the rectification effect of radiofrequency attributable to NLAHE has been demonstrated. NLAHE has also been considered in p-doped trigonal Te crystal [6], as well as a probe for massive Dirac fermion (DF) state because Berry curvature is nonzero only in the gapped state [7].

The system we are considering is an organic conductor $\alpha$-$(BEDT\text{-}TTF)_2I_3$, which is well recognized as a multilayer 2D massless Dirac fermion system with a pair of tilted Dirac cones at high pressure [8]. At the ambient pressure, it shows a metallic conduction at high temperatures. Upon cooling it transitions into an insulating state due to the horizontal-stripe charge ordering (CO) at $T_{CO}$ = 135 K, as schematically shown in Fig. 1. The metallic state belongs to $P\bar{1}$ space group, whereas in its insulating state the CO breaks the inversion symmetry, resulting in $P1$ symmetry and appearance of two domains with different charge distributions [9,10]. With increasing hydrostatic pressure, $T_{CO}$ decreases and, finally, above a critical pressure $P_C \approx 1.3$ GPa the system enters the massless DF state, where the transport shows a nearly temperature-independent, but metallic behavior.

The massless DF state at high pressure in $\alpha$-$(BEDT\text{-}TTF)_2I_3$ has been firmly established. Theoretical work [11–13] show gapless Dirac cones, which are tilted and form at general points $-\mathbf{k_0}$ and $\mathbf{k_0}$ of the 2D Brillouin zone due to its low crystal symmetry [14]. Experimental evidences strongly support the massless DF model. Interlayer magnetoresistance suggests the effect of zero mode Landau level characteristic to DF [15–17]. Specific heat exhibits the characteristic $C \propto T^2$ associated with linear density of states [18]. Furthermore, site-selective NMR supports the tilting of Dirac cones [19].

On the other hand, the details of the transition between the CO state and massless DF remain to be solved. Here, we refer the CO insulator region close to the critical pressure $P_c$ to the weak CO state. It is expected that in the weak CO state the CO is highly suppressed and the Dirac cones are narrowly gapped, realizing insulating massive DF state with finite Berry curvature. Experimentally, the massive DF manifests in the temperature dependent interlayer transport as a reentrant metallic behavior suggesting the edge transport in the insulating state [20]. As for the electronic structure of the weak CO state, however, other possibilities have been proposed. A particular phase diagram for this system suggests a semimetallic CO state between the two states [21]. Furthermore, optical study suggests DFs with a pseudogap behavior at intermediate pressures [22]. More recently, strong evidence for the massive DF in the weak CO state was reported. It has been shown that the temperature-dependent interlayer magnetoresistance features a peak structure, whose magnetic field dependence can only be reproduced when the insulating massive DF picture is taken into consideration [23].

The 2D massive DF system with tilted Dirac cones is the most simple system with finite Berry curvature dipole. Therefore, the weak CO state in $\alpha$-(BEDT-TTF)$_2$I$_3$ provides an ideal platform to investigate Berry curvature dipole effects such as NLAHE. We have previously considered the NLAHE behaviors in the weak CO state of $\alpha$-(BEDT-TTF)$_2$I$_3$ and estimated its magnitude to be in the observable range [24,25]. The present study is aimed to confirm these results experimentally.

## 2. Experimental

A single crystal with a dimension of approximately $0.56 \times 0.48 \times 0.04$ mm$^3$ was used. Eight contacts were attached with carbon paint (Fig. 2(a)), with contact numbering shown in Fig. 2(b). Since a whole crystal was used, the crystallographic axes can be determined from its shape [26]. The line connecting the corners with 99° angle is the direction of the crystallographic *b*-axis. The *a*-axis direction is then almost parallel to a line joining contacts 4-5. The sample was enclosed in a Be-Cu piston-cylinder type pressure cell with Daphne 7373 as the pressure medium. The pressure at room temperature was determined by measuring the resistance of a manganin wire enclosed with the sample. All measurements were performed in the dc limit by using Keithley 2182 nanovoltmeter and Yokogawa 7651 dc current source.

## 3. Result and Discussion

To probe the weak CO state, the sample was pressurized in several steps up to 1.78 GPa. To confirm the applied pressure, we performed temperature dependent resistance measurements as shown in Fig. 2(c). The in-plane resistivity data at 1.25 GPa and resistivity upturn at $T_{\text{Min}}$ = 37 K are consistent with known result [27]. In the CO state, $T_{\text{CO}}$ can be evaluated by finding the peak of $-(d \ln \rho_{zz})/dT$ of the interlayer resistivity. As the weak CO state is approached, the CO transition becomes less abrupt and more second-order like. We evaluated $T_{\text{CO}}$ by finding the peak in $d \ln \rho / d(1/T)$ at each pressure (Fig. 1). At 1.25 GPa, the peak is broad and centered at $T \approx 20$ K. These results justify that at $P$ = 1.25 GPa, the sample is in the desired range of pressure for the weak CO state. Since NLAHE requires carrier imbalance, we checked the carrier density with conventional Hall measurement and determined the majority carrier as holes, with a density of about $n = 10^{17}$ /cm$^3$. We note that the carrier density is one order of magnitude higher than that used in the calculation ($n \sim 10^{16}/\text{cm}^3$) [24,25].

We measured the current dependence of the Hall voltage (the *V-I* curve) from $-1$ mA to 1 mA for four possible current directions at zero magnetic field. To eliminate the mixing of the linear longitudinal resistance due to unexpected but unavoidable misalignment the Hall contacts, we extracted the nonlinear Hall signal by reversing the current direction. The symmetric Hall signal, $V_\perp^S$, from dc *V-I* data can be obtained as

$$V_\perp^S = \frac{V_\perp(I \geq 0) + V_\perp(I \leq 0)}{2}, \quad (1)$$

where $V_\perp$ is the raw signal taken with Hall electrode configuration. However, $V_\perp^S$ is still contaminated by the nonlinear longitudinal resistance component $V_{//}^S$ due to contact offset. To obtain the NLAHE component, $V_{\text{NLAHE}}$, we measured simultaneously signals in Hall ($V_\perp$) and longitudinal ($V_{//}$) configurations. The *V-I* signals were then separated into components symmetric and antisymmetric with respect to current reversal as the following

$$\begin{aligned} V_{//} &= V_{//}^A + V_{//}^S \\ V_\perp &= V_\perp^A + V_\perp^S, \end{aligned} \quad (2)$$

where the labels *S* and *A* denote the symmetric and antisymmetric components, respectively. At zero magnetic field, there should be no voltage component perpendicular to the applied current. Additionally, scattering does not give perpendicular component, except in the case of broken time-reversal symmetry, such as in anomalous Hall effect. Therefore, the voltage which appears as $V_\perp$ must come only from the Hall contact offset and NLAHE if it exists. We then estimated the amount of nonlinear longitudinal voltage component across the Hall electrodes as $kV_{//}$, where *k* is a proportionality factor. With this, the voltage appearing in the Hall configuration can be written as

$$V_\perp = kV_{//}^A + (kV_{//}^S + V_{\text{NLAHE}})$$
$$= V_\perp^A + V_\perp^S. \tag{3}$$

Here, $V_\perp^A$ is zero when the Hall contacts are perfectly aligned. We obtained $V_\perp^S$ following equation (1) and the factor $k$ as $k = V_\perp^A/V_{//}^A$ following equation (3). NLAHE voltage then was obtained by subtracting the longitudinal voltage contribution $kV_{//}^S$ from $V_\perp^S$. The details are provided in supplementary information [28]. From this point, NLAHE signal refers to the symmetrized Hall voltage after the background subtraction.

Symmetrized voltages after background subtraction obtained at $P$ = 1.25 GPa and $T$ = 4.2 K as a function of $I^2$ for various current directions are shown in Fig. 3(a). The symmetrized signals show second order $V \propto I^2$ behavior. Hereafter, we will denote the resistance and voltage as $R_{ij,kl}$ and $V_{ij,kl}$, where the indices $i$, $j$, $k$, and $l$ indicate the positive current, negative current, positive voltage, and negative voltage electrodes, respectively. We also maintain the relative position of the electrodes as shown in the inset of Fig. 3(a) for all measurements. The configuration $V_{45,27}$, where the current is directed almost parallel to the $a$-axis gives the strongest nonlinear signal. The signal weakens as the current direction is rotated towards the $V_{72,45}$ configuration, corresponding to the current direction almost parallel to the $b$-axis. For $V_{45,27}$, $V_{81,36}$, and $V_{72,45}$ the resulting nonlinear voltage is positive. Therefore, the electric fields of the nonlinear voltages for those configurations are directed from the positive voltage probe to the negative voltage probe, as indicated by the solid arrow. However, the nonlinear voltage for the $V_{36,18}$ configuration is negative. Therefore, its nonlinear electric field is directed from the negative to positive voltage probe. NLAHE signal is the strongest when the current is biased parallel to the Dirac cone tilting axis direction. For a 2D system, the optimal current bias direction is constrained only when a mirror symmetry is present. In such a case, the largest signal is produced for current bias perpendicular to the mirror axis [1]. The absence of mirror axis in the CO state of $\alpha$-(BEDT-TTF)$_2$I$_3$ means the Dirac cone tilting direction cannot be determined simply from its crystallographic axis.

The nonlinear current due to the NLAHE in the weak CO state in $\alpha$-(BEDT-TTF)$_2$I$_3$ is represented as, $\mathbf{j}^{(2)} = \left(\chi_{xxy} E_x E_y\right)\mathbf{n}_x + \left(\chi_{yxx} E_x^2\right)\mathbf{n}_y$ [24,25]. Here, the $xy$-plane is parallel to the conducting layer, and the tilting direction of Dirac cones is chosen as the $x$-axis. $\mathbf{E} = (E_x, E_y)$ is the in-plane electric field, and $\mathbf{n}_x$ and $\mathbf{n}_y$ are unit vectors in the $x$- and $y$-directions, respectively. The elements $\chi_{yxx}$ and $\chi_{xxy} = -\chi_{yxx}$ are finite nonlinear Hall conductivity elements, which are represented by the Berry curvature dipole of the system. In the current carrying state, the normal linear transport $\mathbf{j}^{(1)} = \sigma_{xx}\mathbf{E}$ coexists with the NLAHE, where $\sigma_{xx}$ is the scalar longitudinal conductivity. The total current is given by $\mathbf{j} = \mathbf{j}^{(1)} + \mathbf{j}^{(2)}$. Assuming $\left|\left(\chi_{yxx}/\sigma_{xx}^2\right)j\right| \ll 1$, we can obtain the following asymptotic formula for the anomalous Hall field as a function of electric current directed at an angle $\alpha$ away from the $x$-axis (Dirac cone tilting axis) in the presence of normal linear transport

$$E_\perp = -\frac{\chi_{yxx}\cos\alpha}{\sigma_{xx}^3} j^2. \tag{4}$$

Therefore, in the case of simple massive DF systems, symmetrized voltage with a current dependence of the form $V^{(2)} \propto \cos\alpha$ is expected, although it does not hold for general systems with more complicated band dispersion such as WTe$_2$ [29]. In $\alpha$-(BEDT-TTF)$_2$I$_3$, the in-plane resistivity anisotropy is typically within a factor of two [30], so the current directions giving the strongest and the weakest signals should remain perpendicular to each other. The appearance of the strongest signal for the current direction close to the $a$-axis is qualitatively consistent with previous report that the *x*-axis in the weak CO state is rotated by approximately $\theta = -30°$ away from the *a*-axis towards the *b*-axis [20].

We note that the signs of $V_{45, 27}$, $V_{81, 36}$, and $V_{72, 45}$ are consistent with rectification effect since the nonlinear voltages are polarized only in a fixed direction relative to the applied current. For $V_{36, 18}$ the direction is opposite, so it appears to violate the rectification effect. This can be seen in Fig 3(b), where $V_{36, 18}$ and $V_{63, 81}$ deviate from the cosine curve. Although the voltage is second-order so that the reverse current configuration $V_{63, 18}$ will have the same voltage as $V_{36, 18}$ and there will be no issue, as mentioned later, we consider that the measurement of $V_{36, 18}$ has some accidental problem owing to geometrically non-ideal electrode. The angular dependence of the NLAHE voltage can be visualized as Fig. 3(c).

Next, data taken at a higher pressure 1.35 GPa represented by $V_{45, 27}$ and $V_{81, 36}$ are shown in Figs. 3(d) and (e), respectively. The signal drops significantly, almost vanishes. Here, we consider two possibilities. First, relative to the conductivity at 1.25 GPa, there are conductivity increases by about 3 times at 1.35 GPa and by about 5 times at 1.78 GPa (Fig. 2(c)). If we assume a constant Berry curvature dipole at those pressures, this corresponds to one and two order of NLAHE signal magnitude decrease at 1.35 GPa and 1.78 GPa, respectively according to equation (4). If this scenario is true, inversion symmetry must be broken at those pressures in order for NLAHE to exist. In the second scenario, the system has entered the massless DF state at 1.35 GPa. In the massless DF state, the CO transition is completely suppressed; thus inversion symmetry is preserved and second-order conduction is prohibited. This second scenario is more likely considering the data taken at $P = 1.78$ GPa, where the system is known to be in the massless DF state, and consistent with previous reports on the disappearance of $T_{CO}$ in the massless DF state and the interlayer transport [20,23,27,31]. The behavior under pressure thus suggests that the system undergoes a transition into massless DF state at 1.25 GPa < $P_C$ < 1.35 GPa.

For $P$ = 1.25 GPa, the nonlinear Hall coefficient $\chi_{yxx}$ can be estimated using equation (4). Here, we consider $V_{81, 36}$ data because they show a large response, while both current density and longitudinal conductivity can be evaluated with a standard geometry. Further, we limit our analysis for data taken at 4.2 K with an applied current range of 0.5 – 1 mA, where the uncertainty is small. The conductivity at 4.2 K was estimated to be $\sigma_{xx}$ = 19.9 S/cm. The NLAHE electric field $E_\perp$, was taken as the ratio of the symmetrized voltage to the distance between the voltage electrodes (0.41 mm). The current density $j_x$, was estimated from the applied current and the sample geometry. Using the Dirac cone tilting axis direction from ref. [20], we estimated $\chi_{yxx}$ = 0.033 ± 0.004 A/V² at 4.2 K. As a comparison, following ref. [24] and inputting the experimentally obtained carrier density $n$ = 10¹⁷ /cm³ and the charge order gap $\Delta_{CO}$ = $k_B T_{CO}$ with $T_{CO}$ = 20 K, the theoretically predicted value is $\chi_{yxx}$ = 0.36 A/V². The experimental value, therefore, is smaller by one order of magnitude. This discrepancy is discussed again later.

The temperature dependence of the NLAHE signal for $V_{45, 27}$ configuration is shown in Fig. 4. In general, the signal increases with decreasing temperature, particularly at $T$ < 20 K, below $T_{CO}$ since the normal metallic state is inversion symmetric, so second-order behavior is not allowed. The temperature dependent conductivity as a factor $1/\sigma_{xx}^3$, in addition, causes a rapid decrease due to its activated form. The NLAHE signal falls with temperature more rapidly than $1/\sigma_{xx}$, supporting the appearance of the signal at $T$ < 20 K.

The result suggests that NLAHE may exist in $P$ = 1.25 GPa, in the weak CO

state of $\alpha$-(BEDT-TTF)$_2$I$_3$. However, there are several issues still remaining in this NLAHE measurement. So far in other NLAHE experiments [2–5] second order longitudinal voltage is not active, whereas it is present in our experiment. In contrast to them, the present NLAHE was observed in a slightly-doped CO insulator. Considering that the CO state of this system could show nonlinear longitudinal transport [32], the finite second-order longitudinal voltage might be reasonable.

In addition, Hall contact misalignment and their finite size and shapes introduce some ambiguity because it is possible to underestimate or overestimate the correction, which can be the cause of voltage sign problem in $V_{36, 18}$. In this sense, the present result does not necessarily confirm but strongly support the existence of NLAHE in weak CO state in $\alpha$-(BEDT-TTF)$_2$I$_3$.

It should be noted that in a perfect $\alpha$-(BEDT-TTF)$_2$I$_3$, the Fermi energy is located exactly at the Dirac point due to its 3/4 band filling originating from charge transfer between (BEDT-TTF)$^{+1/2}$ and (I$_3$)$^-$. The chemical potential thus falls inside the mass gap in the massive DF state. On the other hand, the theory [1,7] was actually formulated assuming metallic state with Fermi energy, which is the chemical potential at zero temperature, located just outside the gap. In the previous work, on the other hand, we discussed the possible NLAHE in the insulating massive DF system simulating $\alpha$-(BEDT-TTF)$_2$I$_3$ as a function of chemical potential or carrier density imbalance quantitatively [24, 25]. When there exists finite carrier imbalance, the thermally excited carriers cause finite NLAHE at finite temperature even if the chemical potential is located in the CO gap [24]. Experimentally it has been shown that although the low temperature resistivity shows a typical insulating behavior ($d\rho/dT < 0$), infrared measurement results indicate Drude conductivity increase at low frequencies, indicating existence of thermally excited free carriers [22].

Finally, we address the discrepancy between $\chi_{yxx}$ obtained in this experiment and the theory. Here, the data were obtained under zero-field cooling, whereas the theory predicted the necessity of the current-field cooling technique [24,25]. The transition into the CO state breaks the inversion symmetry, resulting in two domains. The two domains are randomly aligned and opposite in sign for NLAHE; thus, under zero field cooling, the signal should vanish. The proposed current-field cooling was expected to align the domains by exploiting dc current-induced magnetization (orbital Edelstein effect), which is also a Berry curvature dipole effect, as the sample is being cooled in a constant magnetic field. As described above, the experimentally determined $\chi_{yxx}$ is an order of magnitude smaller than the predicted value. The discrepancy can be understood since the theoretical prediction assumes a single-domain system, so it gives only the upper limit of $\chi_{yxx}$. The existence of the two types of domains has been investigated by Yamamoto *et al.* using infrared second-harmonic spectroscopy [33]. The domains were found to be mobile and macroscopic in size. Difference in proportion of the domains has been reported by Kakiuchi *et al.* [9], so it is likely that one type of domain already dominates across the sample. Therefore, we assert that the nonzero signal observed in this study resulted from asymmetry in the domain formation.

**4. Conclusions**

In conclusion, we have successfully observed the NLAHE in the weak CO state of $\alpha$-(BEDT-TTF)$_2$I$_3$. The observed NLAHE features, the current direction dependence and the correlation with the CO relation, are consistent with our previous estimation assuming the 2D massive DF with a pair of tilted Dirac cones. This is the first observation of topological transport in organic conductors, and also the first example of NLAHE in the electronic state with spontaneous symmetry breaking. The rather small value of the observed NLAHE can be explained by the cancellation between two types of CO domains.

**Acknowledgements**

The author thanks Dr. T. Taen, Dr. K. Uchida, Ms. A. Mori, Mr. K. Yoshimura, and Dr. M. Sato for valuable discussions and support. This work was supported by JSPS KAKENHI Grant Numbers JP20H01860 and JP21K18594.

\* kiswandhi@issp.u-tokyo.ac.jp

\*\* osada@issp.u-tokyo.ac.jp

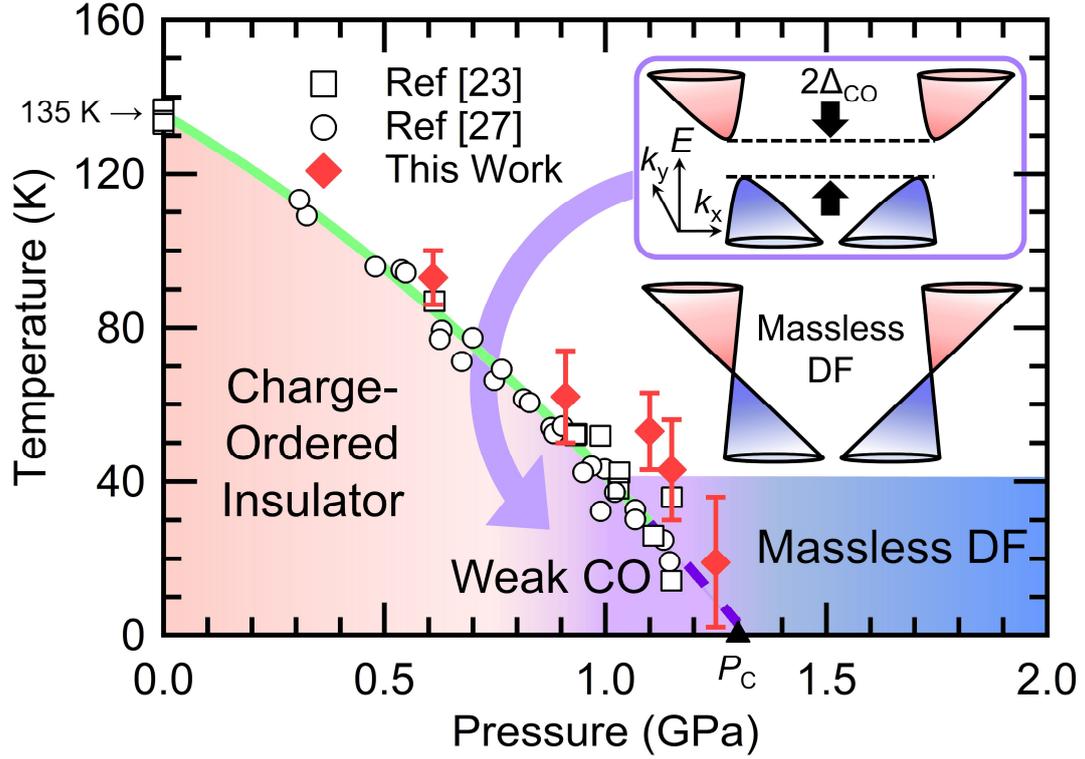

Figure 1 (color online)

Schematic phase diagram of α-(BEDT-TTF)$_2$I$_3$ constructed from resistivity measurements given in refs. [23] and [27]. Red diamonds indicate CO transition temperatures determined in the present work. Weak CO state is expected at transition between CO insulator and massless DF states below the critical pressure, $P_C$. Top inset shows a schematic of the weak CO state with a pair of gapped Dirac cones, realizing massive DF considered in this study. Bottom inset illustrates the massless DF at high pressure.

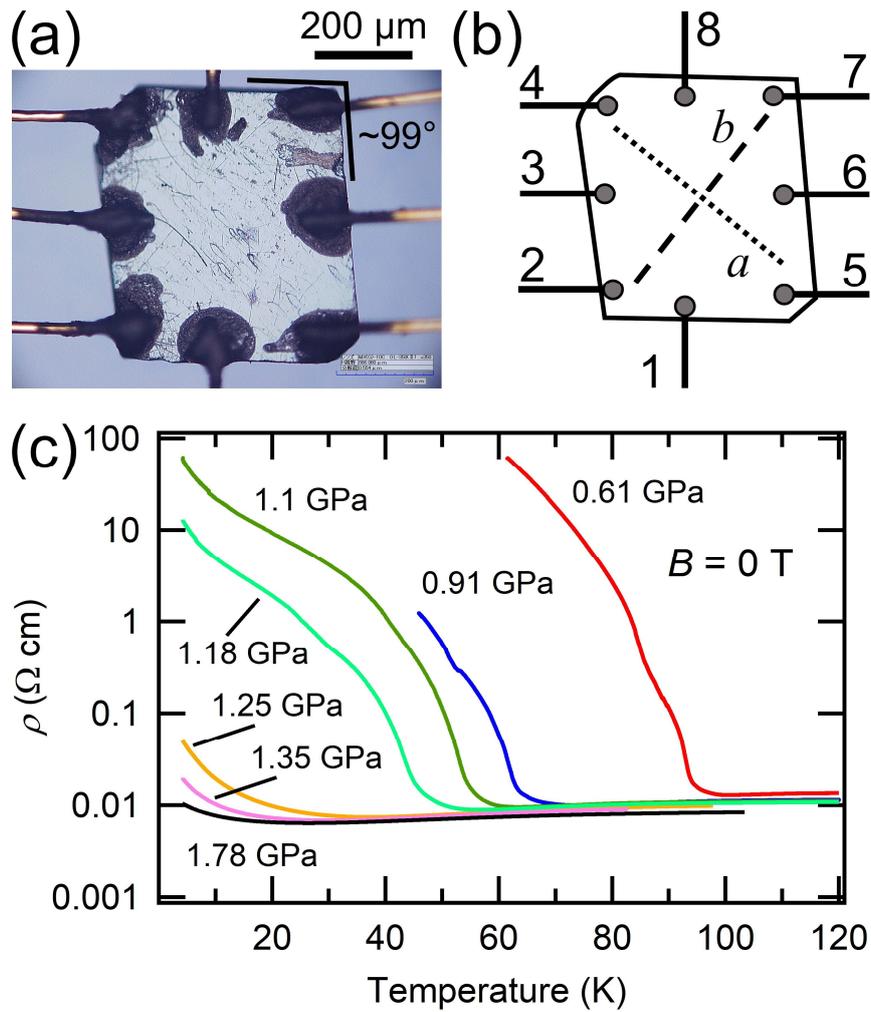

Figure 2 (color online)

Contact configuration and resistivity of $\alpha$-(BEDT-TTF)$_2$I$_3$. (a) Picture of the sample showing the contact configuration. (b) Schematic showing contact numbers used in text and the $a$-axis (dotted line) and the b-axis (broken line) directions inferred from the angle of the sample. (c) Temperature dependent in-plane resistivity data of $\alpha$-(BEDT-TTF)$_2$I$_3$ at various pressures.

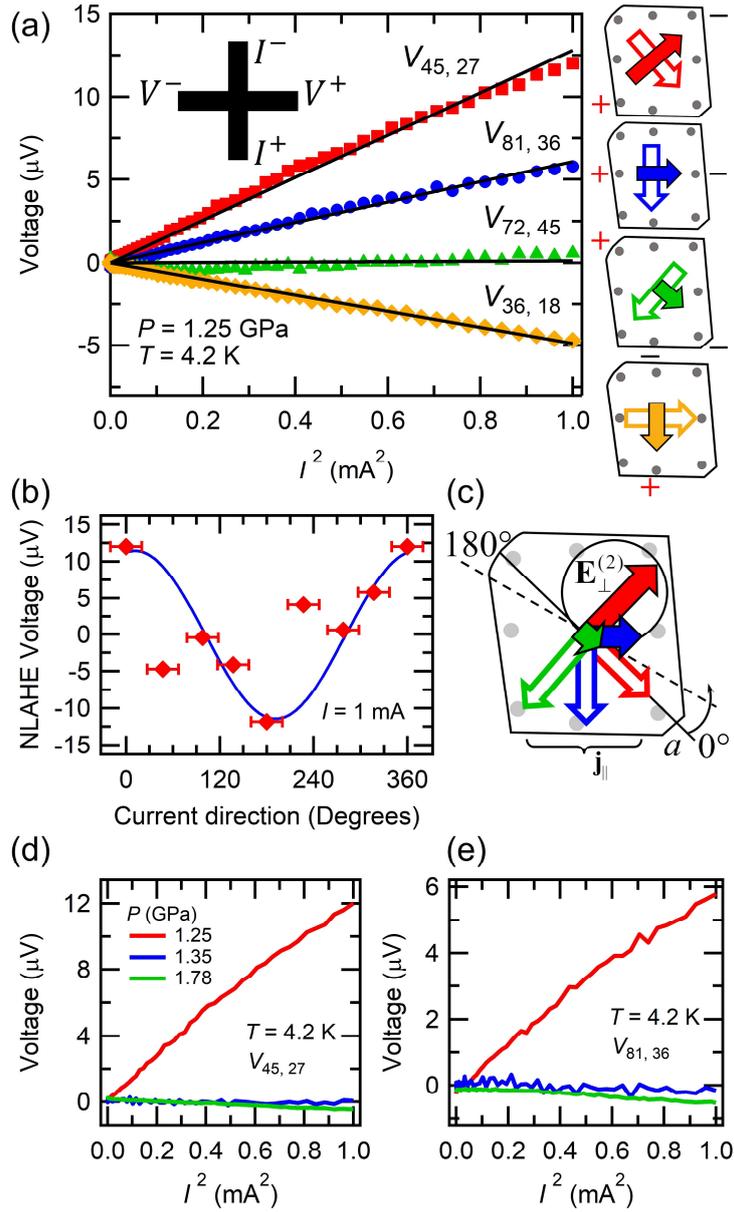

Figure 3 (color online)

NLAHE signal as symmetrized *V-I* curves of zero-field Hall measurement after background subtraction. (a) Signals obtained at various current directions at *P* = 1.25 GPa and *T* = 4.2 K. Inset and figures on the right side indicate the contact configurations. Hollow arrows indicate the positive current direction and the positive and negative signs indicate the voltage probe polarities. Solid arrows indicate the directions of NLAHE electric field, whose magnitude is indicated by the arrow length. (b) Angular dependence of the NLAHE at *I* = 1 mA. The current direction angle is measured counterclockwise relative to the *a*-axis (see panel (c)). The cosine curve is a guide to the eyes. (c) Schematic trajectory of the nonlinear Hall field vector denoted by solid arrows. The circle defines the angular dependence of nonlinear signal strength. The hollow arrows denote the applied current directions. The broken line indicates Dirac cone tilting axis direction. NLAHE signals of (d) $V_{45,\,27}$ and (e) $V_{81,\,36}$ configurations measured at various pressures at $T = 4.2$ K. The signal becomes much smaller at $P = 1.35$ GPa, similar to $P = 1.78$ GPa, which is located in the massless DF regime.

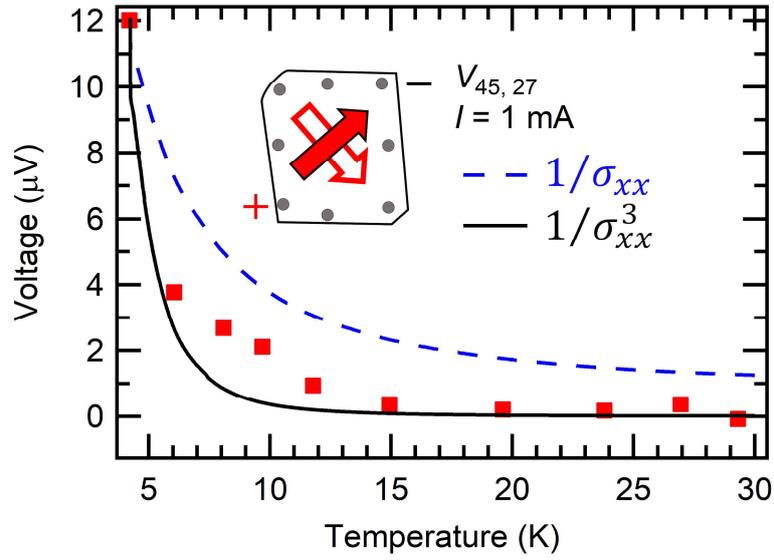

Figure 4 (color online)

Temperature dependence of NLAHE voltage compared with the conductivity $\sigma_{xx}$, represented as $1/\sigma_{xx}$ (broken line) and $1/\sigma_{xx}^3$ (solid line). Here, $\sigma_{xx}$ was taken from longitudinal voltage $V_{45,86}$ normalized to NLAHE voltage at $T$ = 4.2 K.

# Supplementary Information

# Observation of Possible Nonlinear Anomalous Hall Effect in Organic Two-Dimensional Dirac Fermion System

Andhika Kiswandhi* and Toshihito Osada**

*Institute for Solid State Physics, The University of Tokyo,*

*5-1-5 Kashiwanoha, Kashiwa, Chiba 277-8581, Japan.*

**Longitudinal background voltage estimation**

In the main text, we presented the nonlinear anomalous Hall signal after background subtraction. The main issue is that since the Hall contacts were made by hand, there will be some contact offset in the longitudinal direction. This means, the Hall electrodes will capture a portion of the longitudinal voltage. In our measurement, second-order longitudinal voltage exists, so it appears in the nonlinear anomalous Hall measurement. Whereas it is standard to eliminate longitudinal magnetoresistance by magnetic field reversal in conventional Hall effect measurement, NLAHE has no field reversal equivalent. Here, we describe our method of estimating the background voltage.

In our experiment, we measured dc *V-I* curve from −1 to 1mA for both Hall ($V_\perp$) and longitudinal ($V_{//}$) voltages. Generally, signals can be mathematically separated into their symmetric and antisymmetric components. The longitudinal voltage $V_{//}$, can be written as

$$V_{//} = V_{//}^A + V_{//}^S, \qquad (1)$$

where the indices A and S indicate the components antisymmetric and symmetric with respect to current reversal, respectively. When the Hall contacts are offset, $V_\perp$ contains a fraction of the longitudinal signal as the following

$$V_\perp = V_{\text{NLAHE}} + k\left(V_{//}^A + V_{//}^S\right),$$
$$= V_\perp^A + V_\perp^S \qquad (2)$$

where $k$ is a proportionality factor. Here, $V_{\text{NLAHE}}$ and $kV_{//}^S$ are combined into $V_\perp^S$ since it is not possible to distinguish them experimentally. Assuming relation (2) holds, we calculated the proportionality factor $k$ as a function of the applied current from the antisymmetric components as $k = V_\perp^A/V_{//}^A$. With this, the NLAHE voltage then can be obtained as

$$V_{\text{NLAHE}} = V_\perp^S - kV_{//}^S. \qquad (3)$$

As an example, the data taken at $T = 4.2$ K for the $V_{81, 36}$ configuration and the background subtraction are shown in Fig. S1.

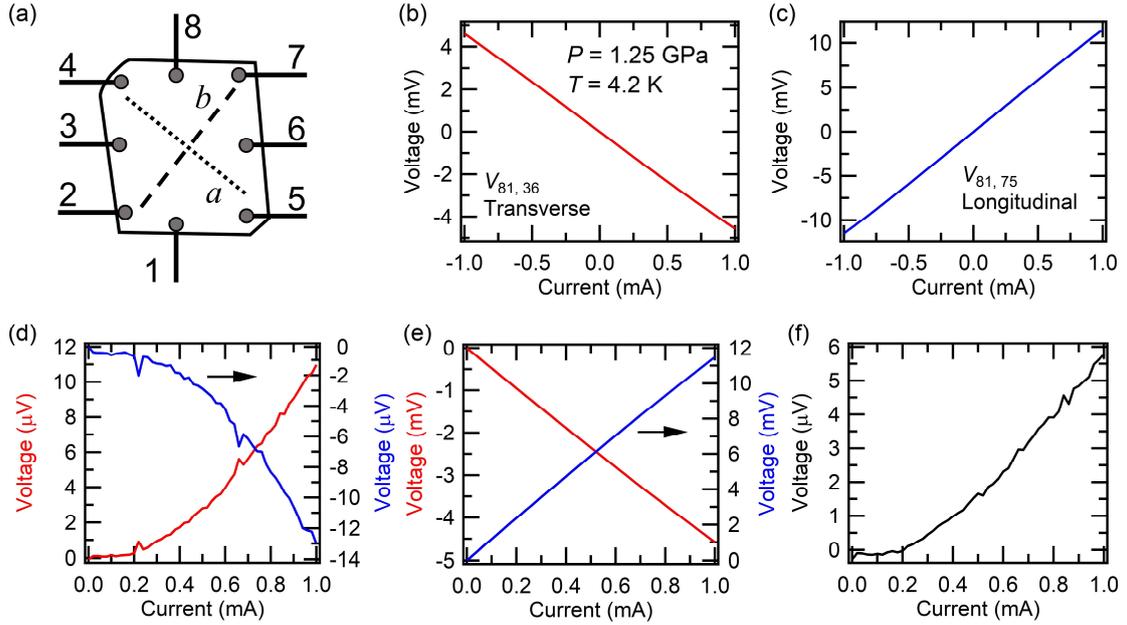

**Figure S1**. Background subtraction procedure using $V_{81, 36}$ data at 1.25 GPa at 4.2 K. (a) Contact configuration, as shown also in Fig. 2(b) as a reference. (b) Raw transverse $V_{81, 36}$ and (c) raw longitudinal $V_{81, 75}$ voltages as collected in experiment. (d) Symmetric part $V_\perp^S$ (red, left axis) and $V_{//}^S$ (blue, right axis) obtained using equation (1). (e) Remaining antisymmetric part of the transverse voltage $V_\perp^A$ (red, left axis) and the longitudinal voltage $V_{//}^A$ (blue, right axis). (f) Transverse $V_{81, 36}$ after the background subtraction.

The signal also drops at pressures higher than 1.25 GPa. The data collected at higher pressures $P = 1.35$ GPa and $P = 1.78$ GPa are shown in Fig. S2 below. The drop in the signal when the pressure is increased is different from the drop in the resistivity (Fig. 2(c)). The drop in the resistivity is about 2.5 times when the pressure is increased from 1.25 GPa to 1.35 GPa. It is roughly 5 times when the pressure is increased from 1.25 GPa to 1.78 GPa. Therefore, it is unlikely that the signal disappearance can be explained by drop in resistivity.

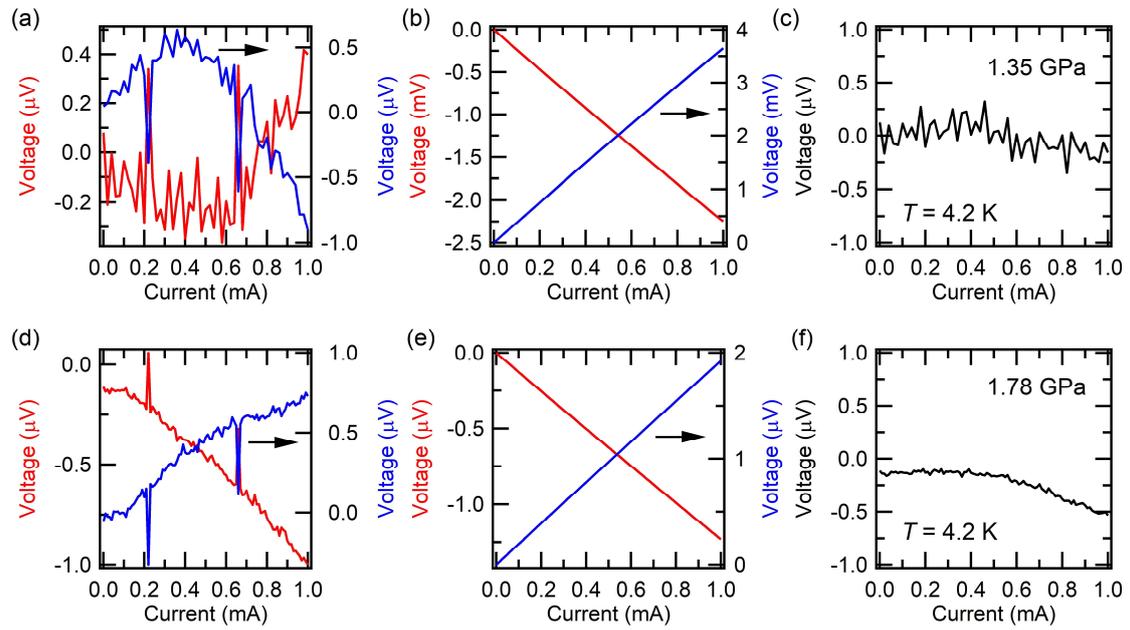

**Figure S2**. Background subtraction of $V_{81, 36}$ data at 1.35 GPa (a-c) and 1.78 GPa (d - f) at 4.2 K. (a) and (d) Symmetric component $V_\perp^S$ (red, left axis) and $V_{//}^S$ (blue, right axis). (b) and (e) Antisymmetric component $V_\perp^A$ (red, left axis) and $V_{//}^A$ (blue, right axis). (c) and (f) The remaining symmetric signal after background subtraction.